\documentclass[12pt]{iopart}
\usepackage{graphicx}

\begin{document}
\title[Replica-symmetry breaking: discrete and continuous
schemes]{Replica-symmetry breaking: discrete and continuous schemes in
  the Sherrington-Kirkpatrick model}

\author{V Jani\v{s}, A Kl\'\i\v{c}, M Ringel }

\address{Institute of Physics, Academy of Sciences of the Czech
  Republic, Na Slovance 2, CZ-18221 Praha, Czech Republic }
\ead{janis@fzu.cz, klic@fzu.cz, ringel@fzu.cz}


\begin{abstract} We study hierarchies of replica-symmetry-breaking
  solutions of the Sherrington-Kirkpatrick model.  Stationarity equations
  for order parameters of solutions with an arbitrary number of
  hierarchies are set and the limit to infinite number of hierarchical
  levels is discussed. In particular, we demonstrate how the
  continuous replica-symmetry breaking scheme of Parisi emerges and
  how the limit to infinite-many hierarchies leads to equations for
  the order-parameter function of the continuous solution. The general
  analysis is accompanied by an explicit asymptotic solution near the
  de Almeida-Thouless instability line in the nonzero magnetic field.
\end{abstract}
 \submitto{JPA} \maketitle

\section{Introduction}\label{sec:Intro}
 
Mean-field theory of spin glasses is now almost complete. It took more
than thirty years from the introduction of a mean-field model by
Sherrington and Kirkpatrick (SK) \cite{Sherrington75} before we
understood its solution and in particular its physical meaning. The
core of the mean-field solution was set rather early by Parisi in his
replica-symmetry breaking (RSB) scheme \cite{Parisi80}.  Parisi
 used, however, the replica trick and a formal procedure
of breaking a symmetry in the (unphysical) replica space when the
limit to zero number of mathematical replicas is performed. Since then
theorists have striven hard to understand the physical meaning of the
Parisi solution and to find alternative ways of its derivation in
order to prove its completeness.

The effort paid off. We now have reached solid understanding of the
physics behind the RSB solution \cite{Parisi07a} and there is a
mathematical proof of exactness of the RSB construction in the SK
model \cite{Guerra03,Talagrand06}. This is not the only output of the
extensive investigation of mean-field spin-glass systems. The
statistical methods developed by studying the infinite-range
spin-glass systems found broad application in interdisciplinary fields
such as informatics, optimisation and computational complexity,
econophysics and biophysics and other frustrated complex and open
systems.

Although our understanding of the mean-field theory of spin glasses is
in global attributes satisfactory, there still remain a few issues
that deserve a more detailed and specific clarification. One of such
questions is the eventual form of the replica-symmetry-breaking solution. Any
derivation of a stable equilibrium spin-glass state uses the so-called
discrete RSB scheme with finite-many hierarchical levels of the order
parameters. Parisi assumed that in the SK model one needs infinite
number of RSB hierarchies and derived an implicit formula for the free
energy of the SK model with a continuous order-parameter function. The
latter is now considered as the exact solution of the SK model.
Unfortunately, the Parisi formula for the averaged free energy of the
SK model is only implicit and no explicit global solution exists.
Moreover, the mathematical proof of exactness of the RSB construction
does not specify whether a discrete or the continuous RSB scheme
produces the maximal free energy. At least in one model, a Potts spin
glass, the one-step RSB solution seems to be stable in a finite region
of temperatures \cite{Gross85}. It is hence important to understand
when one should use the discrete RSB scheme with a finite number of
order parameters and when the continuous limit is appropriate.

The aim of this paper is to analyse properties of the discrete RSB
scheme leading to solutions with a finite number of hierarchical
levels of the order parameters. The emphasis is laid on the way the
discrete scheme goes over to the continuous Parisi solution in the
limit of infinite-many RSB hierarchies. We show that the
continuous limit is a process during which we reduce the degrees
of freedom to a single continuous order-parameter function on a
compact interval. We derive the continuous limit of the stationarity
equations maximising the free energy with large but finite numbers of
RSB hierarchies and thereby we obtain a functional equation for the
order-parameter function from the Parisi solution. We first derive the
equations in the continuous limit generally and then we illustrate the
process of building the continuous limit on the asymptotic expansion
near the de Almeida-Thouless (AT) instability line of the SK model in
the external magnetic field. We explicitly evaluate the leading
asymptotic contribution to the Parisi order-parameter function.

\section{Discrete replica-symmetry breaking scheme}
\label{sec:discrete}

There is no direct way to the Parisi free energy surpassing the
concept of discrete replicas in one or another way. We can either use the
replica trick to handle averaging over random configurations of the spin
coupling \cite{Parisi80} or we can use real replicas with which we test
thermodynamic homogeneity (independence of boundary and initial
conditions) of the chosen macroscopic thermodynamic state \cite{Janis05c}.
 In each of these approaches one must extend the discrete multiplicity of
the replicated phase space to a continuous parameter. This can be done
only for a specific structure of the replicated phase space, namely a
vertical tree containing only child replicas within parental ones.
Communication between replicas of the same generation materialises
exclusively via one or more antecedent generations of replicas (common
ancestors). The phase space forms a hierarchical ultrametric structure. It
is then natural to start the investigation of the RSB construction with the
discrete scheme.

\subsection{Hierarchical free energy and order parameters}
\label{sec:discrete-FE}

A hierarchical character of the phase space of the SK model is
expressed in a hierarchy of order parameters standing for generations
of the replicated spins. Each generation is characterised by a pair of
numbers $m_l$ and $\chi_l$. The former "geometric" parameter expresses
a probability with which the original spins are affected by spins from
the  replicated systems and  the latter represents
 the strength with which the original and the replicated spins  from the
$l$th hierarchy interact \cite{Janis05c}. A phase space with $K$
hierarchies ($K$RSB) is characterised in addition to the SK order
parameter $q$ also by $K$ pairs $\{m_l,\chi_l\}$ with
$l=1,\ldots,K$.\footnote{We use here, in   accordance with
Ref.~\cite{Janis05c}, a decreasing sequence of   overlap susceptibilities
$\chi_l$ instead of parameters $q_l$ in the Parisi notation.  The two
sequences are simply related $\chi_l = q_{K+1-l} - q_0$ for $l=1,\ldots,K
+ 1$ .}   All these parameters are determined from stationarity equations
for a hierarchical free energy with $K$ hierarchies.

It is not the set of pairs $\{m_l, \chi_l\}$ that explicitly appears in
the hierarchical free energy. In fact, we can construct a free energy
functional for either $ \chi_l, \Delta m_l = m_{l-1} - m_l$ or $m_l,
\Delta \chi_l = \chi_{l} - \chi_{l+1}$ for $l=1,\ldots,K$ with boundary
conditions $m_0 =1$ and $ \chi_{K+1} = 0$. Both set of  parameters $m_l$
and $\chi_l$  form a decreasing sequence. The
general formula for the hierarchical free energy with the former pairs was
introduced in Ref.~\cite{Dotsenko01} while the latter in
Ref.~\cite{Janis05c} that we use also in this paper.

The free energy with $K$ hierarchies is characterised by $2K +1$ order
parameters. It is the SK order parameter $q$ and $K$ pairs
$\{m_l,\Delta\chi_l\}$, $l=1,\ldots,K$. The averaged free-energy
density with these order parameters reads
\begin{eqnarray} \label{eq:mf-avfe}\fl
  f^K(q,\Delta\chi_1,\ldots,\Delta  \chi_K; m_1,\ldots,m_K) = - \frac
  1\beta \ln 2 -\frac\beta 4 \left(1-q -\sum_{l=1}^K
    \Delta\chi_l\right)^2\nonumber \\ + \frac \beta 4
\sum_{l=1}^K   m_l\Delta\chi_l\left[2\left(q + \sum_{i=l+1}^{K}\Delta
\chi_i\right)     + \Delta\chi_l\right] - \frac 1\beta
\int_{-\infty}^{\infty}  \mathcal{D}\eta\ \ln \mathcal{Z}_K\ .
\end{eqnarray}
A hierarchical structure of this free energy is evident from the way
its interacting part $\ln\mathcal{Z}_K$ is constructed. It is the
final state in a sequence of partition functions defined inductively
\begin{equation}\label{eq:mf-hierarchy}
  \mathcal{Z}_l = \left[\int_{-\infty}^{\infty}\mathcal{D}\lambda_l\
    \mathcal{Z}_{l-1}^{m_l}\right]^{1/m_l}\ , \end{equation} with the initial
condition $\mathcal{Z}_0 = \cosh\left[ \beta\left(h + \eta\sqrt{q} +
    \sum_{l=1}^K \lambda_l\sqrt{\Delta\chi_l}\right)\right]$. We abbreviated
the Gaussian differential $\mathcal{D}\lambda \equiv d\lambda
\exp\{-\lambda^2/2\}/\sqrt{2\pi}$.

Free energy~\eref{eq:mf-avfe} is a generalisation of 1RSB and 2RSB
solutions obtained by Parisi. It is a generating functional for all
physical quantities of a $K$-level hierarchical solution. The physical
values of the order parameters $q, \Delta\chi_1, m_1,\ldots,\Delta
\chi_K,m_K$ are determined from stationarity equations maximising free
energy $f^K$.  To represent the corresponding stationarity equations
we introduce a set of hierarchical density matrices in the space of
fluctuating random fields $\lambda_l$. We define
$\rho_l(\eta,\lambda_K,\ldots,\lambda_l) = \mathcal{Z}_l^{m_l}/
\left\langle \mathcal{Z}_l^{m_l}\right\rangle_{\lambda_l}$ where
$\langle X(\lambda_l) \rangle_{\lambda_l} =
\int_{-\infty}^{\infty}\mathcal{D}\lambda_l\ X(\lambda_l)$. We further
introduce short-hand notations $t \equiv \tanh\left[\beta\left(h +
    \eta\sqrt{q} + \sum_{l=1}^K \lambda_l\sqrt{\Delta\chi_l}
  \right)\right]$ and $\langle
t\rangle_l(\eta;\lambda_K,\ldots,\lambda_{l+1}) =
\langle\rho_l\ldots\langle\rho_1 t \rangle_{\lambda_1} \ldots
\rangle_{\lambda_l}$.

It is now a straightforward task to derive equations for the order
parameters from the saddle-point equations of functional $f^K$. We
obtain
%
\begin{eqnarray}\label{eq:mfeqs-q}
  q &= \langle\langle t\rangle_K^2\rangle_\eta\ ,\\ \label{eq:mfeqs:Delta_chi}
  \Delta\chi_l &= \langle\langle \langle t\rangle_{l-1}^2
  \rangle_K\rangle_\eta - \langle\langle\langle t\rangle_{l}^2
  \rangle_K\rangle_\eta\ , \\
  \label{eq:mfeqs-geometric}
  m_l \Delta\chi_l &= \frac 4{\beta^2}\ \frac {\langle\langle \ln
    Z_{l-1} \rangle_K\rangle_\eta - \langle\langle \ln Z_{l}
    \rangle_K\rangle_\eta } {2\left(q +
      \sum_{i=l+1}^{K}\Delta\chi_i\right)
    + \Delta\chi_l}\ \end{eqnarray} 
where index $l=1,\ldots,K$.

The discrete RSB scheme does not determine a single solution of the
original spin model, but rather a set of solutions labelled by the
number of hierarchies explicitly taken into account. Parameter $K$ is
hence a free index that is not determined from the free energy. Its
physical value is fixed by thermodynamic stability. We take
so many hierarchies of replicas into account till we reach a stable or
marginally stable solution. Stable solutions with $K$ hierarchies obey
$K+1$ stability conditions. They are a generalisation of the de
Almeida-Thouless stability condition of the replica-symmetric ($K=0$)
solution \cite{Almeida78}. They reflect non-negativity of eigenvalues
of the spin-glass susceptibility \cite{Janis06a} and read
\numparts\label{eq:AT}
\begin{equation}\label{eq:AT-hierarchical}
  \Lambda_K(l) =  1 - \beta^2\left\langle\left\langle \left\langle 1 -
        t^2 + \sum_{i=1}^{l} m_i \left(\langle
          t\rangle_{i-1}^2 - \langle t\rangle_i^2\right)\right\rangle_{l}^2
    \right\rangle_K\right\rangle_\eta \ge 0\ .
\end{equation}
and
\begin{equation}\label{eq:AT-zero}
  \Lambda_K(0) = 1 - \beta^2\left\langle\left\langle \left( 1 -
        t^2 \right)^2  \right\rangle_K\right\rangle_\eta \ge 0\ .
\end{equation}\endnumparts%

With the above equations we are equipped with all the necessary tools
for finding a solution of the SK model.  In particular, we can decide
whether only a finite number of replica generations is sufficient to
reach a stable or marginally stable solution or whether we must go to
 infinite-many replicas and the Parisi continuous limit. Parisi
deduced from 1RSB and 2RSB solutions and confirmed by a truncated
model that indeed we need infinite number of replica
hierarchies. We have recently confirmed this conclusion by solving exactly
stationarity equations \eref{eq:mfeqs-q}-\eref{eq:mfeqs-geometric} in
the asymptotic region near the critical point in zero magnetic field
\cite{Janis06b}. We now extend this asymptotic solution by involving
the non-zero external magnetic field.

\subsection{Asymptotic solution near de Almeida-Thouless instability
  line}
\label{sec:discrete:asymptotic}

Proximity of the instability line naturally introduces a small
parameter that we use in an expansion of
equations~\eref{eq:mfeqs-q}-\eref{eq:mfeqs-geometric} for the order
parameters. If we denote $t_0 = \tanh [\beta (h + \eta\sqrt{q_0})]$,
where $q_0=\langle t_0^2\rangle_\eta$ is the SK order parameter in the
replica symmetric solution, we can define the small parameter to be
\begin{equation}\label{eq:alpha-def}
  \alpha = \beta^2\left\langle (1 - t_0^2)^2\right\rangle_\eta  - 1 > 0\ .
\end{equation}
It measures the distance from the AT line as well as a deviation from
the SK solution. Since the magnetic field is nonzero, only $\chi_l$
for $l=1,\ldots,K$ and $\Delta m_l$ for $l=2,\ldots,K$ are small. The
parameters $q$ and $m_l$ are not small unlike the case of zero
magnetic field \cite{Janis06b}. We determine the dominant asymptotic
behaviour of the corrections to the SK solution in the small parameter
$\alpha$ for an arbitrary number of hierarchical levels $K$.

To derive the leading asymptotic behaviour we must expand
Eqs.~\eref{eq:mfeqs-q} - \eref{eq:mfeqs-geometric} to first two
nontrivial orders of the small parameter $\alpha$. We first need to know
the two leading asymptotic orders of the parameter $q_{EA} = q +
\sum_{l=1}^K \Delta  \chi_l = q +  \chi_1$ to be able to determine the
leading asymptotic behaviour of $\Delta\chi_l$ and $\Delta m_l$ for $l
> 1$.  It means, if we want to go beyond the one-step RSB solution.

The asymptotic limit of the RSB solutions with finite numbers of
hierarchies practically amounts to an expansion in powers of
differences $\Delta\chi_l$. Each integral over the random field
$\lambda_l$ must be expanded at least up to $\Delta\chi_l^3$ to
determine the leading asymptotic behaviour of the order parameters. One
can rather easily calculate the leading orders of $ \chi_1$ and
$m_l$. These two parameters do not depend in the leading asymptotic
order on the number of hierarchies used and are determined from
$1$RSB. To see the dependence of the order parameters on $K$ means to
generate separate equations for single $\Delta\chi_l$ with $l >1$. We
must, however, lift up a degeneracy in the stationarity equations and
expand them up to $\Delta\chi_l^4$. It is a rather tedious task and we
accomplished it with the aid of the programme MATHEMATICA. The
expansion proceeds in the same manner we presented in
Ref.~\cite{Janis06b}. We hence do not repeat the detailed steps of the
expansion but rather summarise the principal findings.

\begin{figure}\centerline{
    \includegraphics[width=9cm]{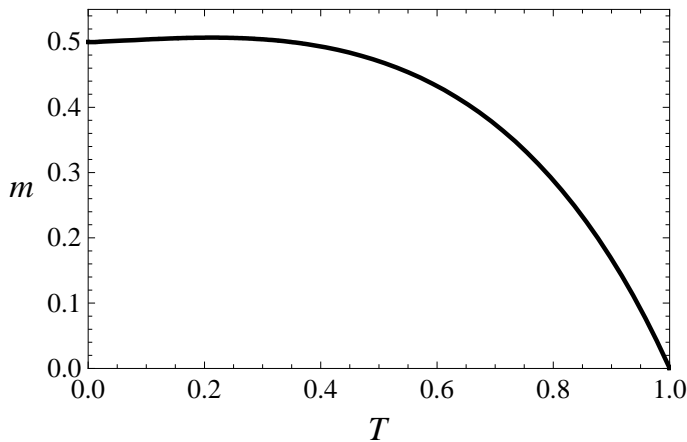} }
  \caption{\label{fig:m1} The limiting value of parameter $m$ from
$1$RSB at the AT instability line as a function of temperature.     The
temperature scale was chosen so that $T_c(h=0)     =1$. } \end{figure}

We must first expand the SK parameter $q$ to the two lowest nontrivial
orders in $\alpha$. The solution is then used to determine the lowest
asymptotic order of $ \chi_1$ and $m_l$.  We obtain $m_l = m +
O(\alpha)$ with
\begin{equation}\label{eq:m0-solution}
  m =  \frac {2\langle t_0^2(1 - t_0^2)^2\rangle_{\eta}}{\langle (1
    -  t_0^2)^3\rangle_{\eta}}
\end{equation}
and
\begin{eqnarray}\label{eq:X-solution}
  \chi_1 = \frac{\alpha}{2\beta^2m} \ \frac 1{1 -
      3\beta^2\left\langle t_0^2(1 -  t_0^2)^2\right\rangle_\eta} +
O(\alpha^2)\  \ . \end{eqnarray}
These two parameters do not depend on the number of hierarchical
levels used. We remind that $ \chi_1 = q_{EA} - q$. The values of the
temperature and the magnetic field are taken from the SK solution.
Parameter $m$ is of order unity even at the boundary of the
spin-glass phase (AT line) where the small parameter $\alpha$
vanishes. Its temperature dependence at the AT line is plotted in
\Fref{fig:m1}.

\begin{figure}\centerline{
   \includegraphics[width=9cm]{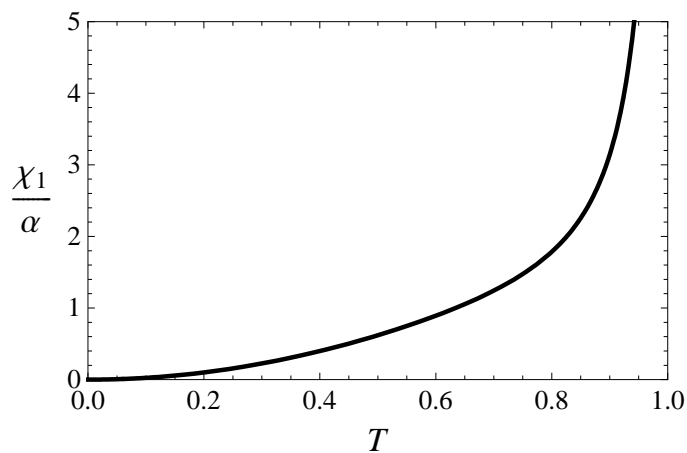}} \caption{\label{fig:X}Proportionality
   of the order parameter $ \chi_1$ to the small parameter $\alpha$
   along the AT line. The ratio diverges at the critical
   temperature $T_c=1$ as $(T_c - T)^{-1}$ and vanishes at zero
   temperature as $T^2$.} \end{figure}

The other parameter of the $1$RSB solution, $ \chi_1$, is proportional
to the small parameter $\alpha$ from Eq.~\eref{eq:alpha-def} and
vanishes at the boundary of the spin-glass phase. Its ratio
$ \chi_1/\alpha$ at the AT line as a function of temperature is
plotted in \Fref{fig:X}.
The ratio diverges in zero magnetic field where $m=0$ and both
$ \chi_1$ and $m$ are linearly proportional to $\theta = (T_c -
T)/T_c$, while $\alpha\sim 2\theta^2$ \cite{Janis06b}.

It is interesting to notice that the asymptotic solution near the AT line
reduces to $1$RSB in non-zero magnetic fields. The SK solution can be
asymptotically correct in the leading order only in zero field where the
parameter $m_1 = 0$. In non-zero fields one has to go to $1$RSB even in
the lowest asymptotic order below the instability line. The existence of
the AT line hence indicates a replica-symmetry breaking. The instability in
the magnetic field does not specify whether the discrete or the continuous
RSB scheme applies in the low-temperature phase. Notice also that
condition $ \chi_1 > 0$ does not necessarily indicate a deviation from the
SK solution. If $m_1 = 0$ then $q_{EA} = q + \chi_1 = q_{SK}$. Only if
both parameters $ \chi_1$ and $m_1$ are simultaneously positive the
physics of the SK solution is changed to $1$RSB.

To disclose the leading asymptotic behaviour of each separate parameter
$\Delta\chi_l$ and $\Delta m_l$ for $l=1,\ldots, K$ we must go beyond
$1$RSB and the leading orders in parameters $m$ and $ \chi_1$. It is
firstly the fourth order in $\alpha$ in Eq.~\eref{eq:mfeqs-geometric}
from which we find that $\Delta\chi_l \doteq \chi_1^1/K$ and
\numparts
\begin{eqnarray}\label{eq:ml}
m^K_l \doteq m_1^1 + \frac {K + 1 - 2l}K\ \Delta m
\end{eqnarray}
where we added a superscript to specify the number of hierarchical levels
used to determine the order parameters $\chi_l,m_l$. Further on, we
introduced a parameter independent of the number of hierarchies  $\Delta m
=m_1^2 - m_2^2$.  This parameter has an explicit asymptotic representation
\begin{eqnarray}\label{eq:M-solution}\fl\qquad \Delta m \doteq \frac
  {\beta ^2
    \chi_1\left\langle \left(1 - t_0^2\right)^2 \left(2 \left(1-3
          t_0^2\right)^2+3 \left(t_0^2-1\right) m \left(8
          t_0^2+\left(t_0^2-1\right) m \right)\right)
    \right\rangle_\eta}{\left\langle (1 - t_0^2)^3 \right\rangle_\eta}
\end{eqnarray}\endnumparts

\begin{figure}\centerline{
 \includegraphics[width=9cm]{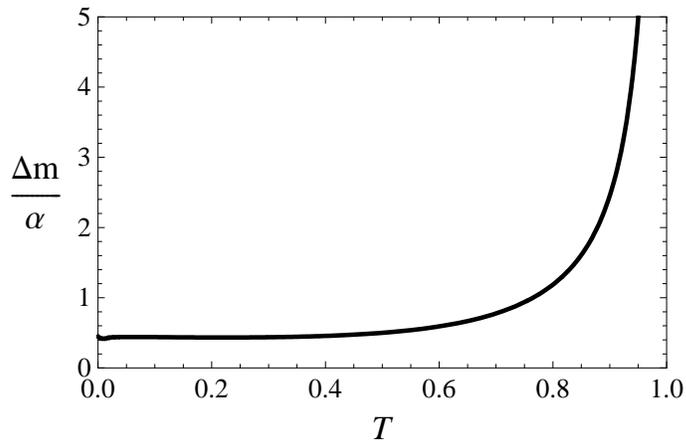} }  \caption{\label{fig:M2}The
 leading-order difference $\Delta m$  factorised by the
 small parameter $\alpha$ to make it of order unity along the AT
 line. It diverges at the critical temperature $T_c=1$ as $(T_c -
 T)^{-1}$.} \end{figure}

Both parameters $ \chi_1$  and $\Delta m$ are linearly proportional to
$\alpha$. The former, however, exists already in $1$RSB, while the latter
first emerges  in  $2$RSB. Since they do not depend on the number of
hierarchies used and determines a uniform distribution of parameters $m_l$
for $l = 3,\ldots,K$, we demonstrated that all characteristic features of
the asymptotic solution near the AT instability line are contained already
in $2$RSB. What was, however, highly nontrivial was to unveil equidistant
distributions of both parameters $ \chi_l$ and $ m_l$. Temperature
dependence of the ratio $\Delta m/ \alpha$ along the AT line is plotted in
Fig.~\ref{fig:M2}.

Finally we evaluated the instability conditions from
Eqs.~\eref{eq:AT-hierarchical} and~\eref{eq:AT-zero}.  They all coincide in
the leading asymptotic order in $\alpha$. We find that all the discrete RSB
solutions are unstable. Instability of the discrete scheme is measured by
the small parameter $\alpha$. The SK ($K=0$) solution has the instability
expressed via the AT condition
\numparts
\begin{equation}\label{eq:Lambda_0}
  \Lambda_0 = 1 - \beta^2\left\langle (1 - t_0^2)^2\right\rangle_\eta =
  -\alpha\ .
\end{equation}
The RSB solutions ($K\ge 1$) improve upon stability of the SK solution
in that their instability is proportional to $\alpha^2$.  We
derived the following explicit expression
\begin{equation}\label{eq:Lambda_K}
  \Lambda_K = - \ \frac{2\beta^2}{3 K^2} \ \frac{ \chi_1 \Delta m}{m +
    2 }\end{equation} \endnumparts
where $\Delta m$ was defined in Eq.~\eref{eq:M-solution}.  The
difference in the order of magnitude in the instability of the SK and
RSB solutions is caused by the existence of the non-zero parameter $m$
at the AT line where $ \chi_l =0$. It is hence impossible for the SK
solution without $m$ to reproduce the exact solution in the non-zero
magnetic field even asymptotically with $\alpha\to 0$.

\begin{figure}\centerline{
 \includegraphics[width=9cm]{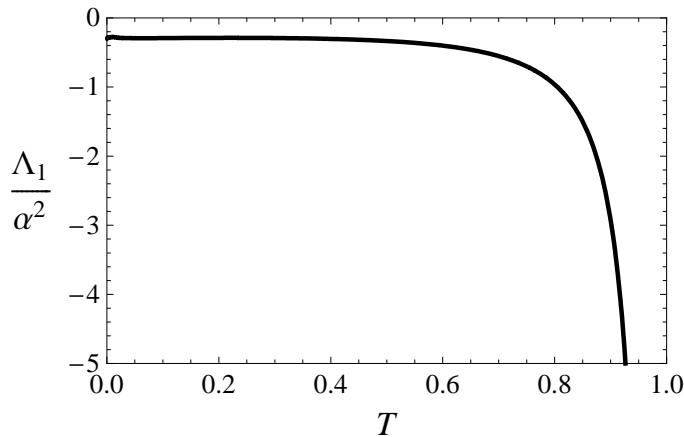}
} \caption{\label{fig:Lambda} Leading asymptotic
  contribution to the instability of 1RSB at the AT line. It diverges at
  the critical temperature as $(T_c - T)^{-2}$.} \end{figure}

The stability conditions of the discrete RSB solutions manifest that
only the continuous limit with $K\to \infty$ becomes marginally
stable. The instability of $1$RSB is plotted in \Fref
{fig:Lambda}. The leading-order term diverges at the critical
temperature $T_c=1$, since the instabilities of the SK as well as of the
discrete RSB solutions in zero magnetic field are proportional to
$\alpha\propto (T_c - T)^2/T_c^2 $ \cite{Janis06b}.

\section{Continuous replica symmetry breaking}
\label{sec:continuous}

The above asymptotic solution corroborates the conclusion of earlier
calculations in zero magnetic field that to reach a stable and
consistent solution we need infinite-many hierarchical levels in free
energy~\eref{eq:mf-avfe}. It is not generally guaranteed that the
limit to infinite hierarchies of the discrete RSB scheme must lead to
the continuous solution.  The distribution of parameters $\Delta\chi_l
=  \chi_1/K$ calculated in the preceding section explicitly manifests
that the Parisi solution is the only possible marginally stable
solution of the SK model in the external magnetic field near the AT
line. Moreover, one of us has recently proved that the Parisi solution
with a continuous order-parameter function can always be constructed
and is marginally stable in the entire spin-glass phase
\cite{Janis07c}. Here we show that the stationarity equations of the
solution from Ref.~\cite{Janis07c} result from the continuous limit of
stationarity equations~\eref{eq:mfeqs-q}-\eref{eq:mfeqs-geometric} of
the discrete scheme.

\subsection{Homogeneous limit to infinite number of hierarchies}
\label{sec:continuous-limit}

The basic assumption of the continuous limit of the discrete RSB
scheme with the number of hierarchies $K\to\infty$ is a uniform
distribution of the differences $\Delta\chi_l$. That is, they are
independent of $l$ and are proportional to $K^{-1}$. We then can
introduce a differential $d x = \lim_{K\to\infty}  \chi_1/K$. It
is not, however, necessary that all parameters $\Delta\chi_l$ are
equal in the asymptotic limit $K\to\infty$ to end up in a continuous
theory.  The differences may vary by factors of order unity.

It is convenient to map the hierarchy indices $l$ on interval $[0,1]$
by introducing a continuous variable $x= \lim_{K\to\infty}( K -  l)/K$.
The continuous version of the hierarchical free energy results from a
process in which we systematically neglect all higher than linear orders
of the differential $dx$ \cite{Duplantier81}. It means that in the
continuous limit we take into account only the second moments of the
Gaussian integrations over the auxiliary fields $\lambda_l$. We first
apply this procedure to the interacting free energy~\eref{eq:mf-avfe} and
then to stationarity
equations~\eref{eq:mfeqs-q}-\eref{eq:mfeqs-geometric}.

We denote $g_l \equiv \ln \mathcal{Z}_l$. Using the notation from the
preceding section we obtain by cutting the expansion of Eq.~\eref
{eq:mf-hierarchy} at the order $O(\Delta\chi_l)$
\begin{eqnarray}\label{eq:g_l-calcul}\fl g_l = \ln
  \left\langle\mathcal{Z}_{l-1}^{m_{l-1}}\right\rangle_{\lambda_l}^{1/m_{l-1}}
  \to \frac 1{m_{l-1}} \ln \left\{\mathcal{Z}_{l-1}^{m_{l-1}}  \left[ 1
    \right.\right. \nonumber\\ \left. \left. + \frac {m_{l-1}}2 \Delta\chi_l
      \left(g_{l-1}'' + m_{l-1}g_{l-1}^{\prime 2}\right) \right] \right\} =
  g_{l-1} + \frac{\Delta  \chi_l}2 \left(g_{l-1}'' + m_{l-1}g_{l-1}^{\prime
      2}\right)\ . \end{eqnarray}
We denoted $g_l' \equiv \partial g_l/\partial h$. The derivatives with
respect to the magnetic field stand for the action of the fluctuating
field $\lambda_l$ when only the second moment contributes to the integral.
In the continuous limit we obtain the Parisi differential equation by
replacing the hierarchy index $l$ by the continuous variable $x$
\begin{equation}\label{eq:g-continuous}
  \frac{\partial g(x, h)}{\partial x} =  \frac {\dot{ \chi}(x)}{2}
  \left[\frac{\partial^2 g(x, h)}{\partial h^2} + m(x)
    \left(\frac{\partial g(x, h)} {\partial h} \right)^2 \right]\ .
\end{equation}
We denoted $\dot{ \chi}(x) \equiv d \chi(x)/d x$. The right-hand side of
Eq.~\eref{eq:g-continuous} has opposite sign to the original Parisi
equation, which is caused by a different assignment of the continuous variable
$x$ to the hierarchy index. Parisi used $x^P = \lim_{K\to\infty}\
\ l/ K$.

To derive the continuous limit of the stationarity equations we must
first find a reduction of the density matrix $\rho_l$ in the limit of
infinite hierarchical levels. Since only a linear term in
$\Delta\chi_l$ contributes and the density matrix is normalised to
unity we have
\begin{eqnarray}\label{eq:rho-def}
  \rho_l \to 1 + m_{l-1} g'_{l-1}\Delta\chi_l \frac {\partial}{\partial h}\ .
\end{eqnarray}
The operator of the derivative with respect to the magnetic field
stands for the random variable $\lambda_l$. Integrals over this random
variable of functions $f(\lambda_l,h)$ weighted with the density
matrix $\rho_l$  reduce in the continuous limit to
\begin{eqnarray}\label{eq:rho-integral}\fl\qquad
  \left\langle \rho_l f(\lambda_l, h) \right\rangle_{\lambda_l} = f(0, h)  +
  m_{l-1} \Delta\chi_l \ g'_{l-1} \frac {\partial f(0,h)}{\partial h} + \frac
  {\Delta\chi_l}{2} \frac{\partial^2 f(0, h)}{\partial h^2}\ .
\end{eqnarray}

We need to evaluate multiple integrals over a number of random fields
$\lambda_l$. We denote
\numparts
\begin{eqnarray}\label{eq:rho-multiple-def}
  f_{l,i} = \left\langle \rho_{l+i} \ldots\left\langle \rho_{l+1}
      f\right\rangle_{\lambda_{l+1}}\ldots \right\rangle_{\lambda_{l+i}}
\end{eqnarray}
and using rule~\eref{eq:rho-integral} we end up with
\begin{eqnarray}\label{eq:rho-multiple-integral}\fl
  f_{l, i+1} = \left\langle \rho_{l+i+1} f_{l,
      i}\right\rangle_{\lambda_{l+i+1}} = f_{l, i} + \Delta\chi_{l+i+1} \left[
    m_{l+i} g'_{l+i} f'_{l, i} + \frac 12 f''_{l, i}\right]\ .
\end{eqnarray}\endnumparts
The increment in the second index can again be represented in the
continuous limit via a differential equation
\begin{equation}\label{eq:f_l-continuous}
  \frac{\partial f_l(X, h)}{\partial X} =  \dot{ \chi}(X)
  \left[ m_l(X) \frac{\partial g(X, h)} {\partial h} \frac{\partial f_l(X,
      h)}{\partial h}   + \frac 12 \frac{\partial^2 f_l(X, h)}{\partial
      h^2}\right] \ . \end{equation}
The solution of the above equation  can be  represented in form of a
"time-ordered" exponential  with
differential operators \cite{Janis07c}
\begin{equation}\label{eq:f_l-integral}\fl
  f_x(X, h) =   \mbox{T}_y \exp\left\{\int_x^X\!\! d y\ \dot{\chi}(y)
    \left[\frac 12 \partial^2_{\bar{h}} + m(y) g'(y, h +
      \bar{h})\partial_{\bar{h}}  \right] \right\} f_x(0, h+
  \bar{h})\bigg|_{\bar{h} = 0}\ . \end{equation}
The ordering operator $T_y$
\begin{equation}\label{eq:T-product}\fl \qquad
  T_y \exp\left\{\int_a^b d y \widehat{O}(y)\right\} \equiv
  1 + \sum_{n=1}^\infty \int_a^b d y_1
  \int_a^{y_1}d y_2\ldots\int_0^{y_{n-1}}\!\! d y_n
  \widehat{O}(y_1)\ldots \widehat{O}(y_n) \end{equation}
orders products of $y$-dependent non-commuting operators from left to
right in a $y$-decreasing succession. It is a standard tool used in
many-body quantum theory to represent time-dependent perturbation
expansion. It is easy to check that function $f_x(X, h)$ obeys
Eq.~\eref{eq:f_l-continuous}.

With the aid of solution~\eref{eq:f_l-integral} we can represent any
physical quantity in the continuous limit. First among them are the
equations for the order parameters.

\subsection{Stationarity equations and stability conditions}
\label{sec:continuous-equations}

To derive an equation for the SK order parameter $q$ we simply put
$f_x(0,h) = t (h) \equiv \tanh(\beta h)$. We obtain from
Eq.~\eref{eq:f_l-integral}
\begin{equation}\label{eq:--x}\fl \qquad
  t(X, h) =   \mbox{T}_y \exp\left\{\int_0^X d y\ \dot{\chi}(y)
\left[\frac
      12 \partial^2_{\bar{h}} + m(y) g'(y, h + \bar{h})\partial_{\bar{h}}  \right]
  \right\} t(h+ \bar{h})\bigg|_{\bar{h} = 0}\ .
\end{equation}
Using this representation in Eq.~\eref{eq:mfeqs-q} we reach at a
generalisation of the SK relation
\begin{equation}\label{eq:q-eq}
  q =  \left\langle t(1, h +
    \eta\sqrt{q})^2\right\rangle_\eta\ . \end{equation}

To evaluate the right-hand side of Eq.~\eref{eq:mfeqs:Delta_chi} we
realise that
\begin{equation*} \left\langle \rho_l \left\langle
      t\right\rangle_{l-1}^2\right\rangle_{\lambda_l} - \left\langle
    \rho_l \left\langle
      t\right\rangle_{l-1}\right\rangle_{\lambda_l}^2 \to \Delta\chi_l
  \left\langle t \right\rangle^{\prime\ 2}_{l-1}
\end{equation*}
from which we find with the aid of integral
representation~\eref{eq:f_l-integral}
\begin{eqnarray}\label{eq:chi-eq}
  \dot{ \chi}(x) &  = \dot{ \chi}(x) \left\langle \mbox{T}_y
    \exp\left\{\int_x^1 d y \ \dot{ \chi}(y) \left[\frac 12
        \partial^2_{\bar{h}} + m(y) g'(y, h + \bar{h})\partial_{\bar{h}}  \right]
    \right\}\right. \nonumber \\  & \left. \qquad \times  (\partial_{\bar{h}}
    t(x, h_\eta + \bar{h}))^2\bigg|_{\bar{h} = 0}\right\rangle_\eta \ .
\end{eqnarray}
We denoted $h_\eta = h + \eta\sqrt{q}$. Equation~\eref{eq:chi-eq} is
fulfilled for all $x\in [0,1]$. It essentially determines the
functional dependence $\dot{\chi}(x)$. We know from the discrete
scheme that $\dot{\chi}(x) >0$.
 
The last relation to be rewritten in the continuous limit is
equation~\eref{eq:mfeqs-geometric}. It is not difficult to reach a
representation
\begin{eqnarray}\label{eq:m-eq}\fl
  m(x)(q +  \chi(1) -  \chi(x)) \nonumber\\ \fl\quad  = m(x) \left\langle
    \mbox{T}_y \exp\left\{\int_x^1 d y\ \dot{ \chi}(y)  \left[\frac 12 \partial^2_{\bar{h}} +
        m(y) g'(y, h + \bar{h})\partial_{\bar{h}}  \right] \right\} t(x, h_\eta +
    \bar{h})^2\bigg|_{\bar{h} = 0}\right\rangle_\eta \end{eqnarray}
that now determines the functional dependence $m(x)$. Again from the
discrete RSB scheme we know that $\dot{m}(x) < 0$.  The two functional
equations~\eref{eq:chi-eq} and~\eref{eq:m-eq} allow for a trivial
solution reducing thereby the RSB quantities to the SK ones.

The derived equations for the continuous version of the order
parameters from the discrete RSB scheme enable us to understand how we
get rid of one functional order parameter. Namely, the function
$\dot{\chi}(x)$ appears in all physical quantities only under
integrals over the index variable $x$. We hence can redefine the
differential $dx \to d \chi = dx\ \dot{\chi}(x)$, since $\dot{ \chi}(x) >
0$. We do not need to know the point-wise dependence $\dot{ \chi}(x)$
to determine physical properties of the low-temperature spin-glass
state. We hence can transform the defining interval $x\in [0,1]$ to a
new one $ \chi\in [0,X]$, where $X = \chi(1)\le 1$. The largest value of
$ \chi(x)$ is the only parameter we need to know from this function. It
must be determined from a stationarity equation and is related to the
Edwards-Anderson parameter by an equation $q_{EA} = q + X$. When we
resign on the explicit dependence $\dot{\chi}(x)$ we also have to
disregard Eq.~\eref{eq:chi-eq}. Only stationarity equations for $q$
and $m( \chi)$, Eqs.~\eref{eq:q-eq} and~\eref{eq:m-eq}, respectively,
remain then relevant. They coincide with the equations derived from
stationarity conditions imposed upon the Parisi free energy in
Ref.~\cite{Janis07c}.

Last but not least we have to find the continuous version of stability
conditions~\eref{eq:AT-hierarchical} and~\eref{eq:AT-zero}. It is
again straightforward to use the above results and integral
representation~\eref{eq:f_l-integral} to arrive at
\begin{eqnarray}\label{eq:stability-continuous}\fl
 \Lambda(x)  =&\fl\ \qquad\quad  1 -  \beta^2 \left\langle \mbox{T}_y
\exp\left\{\int_x^1 d
      y\ \dot{ \chi}(y) \left[\frac 12 \partial^2_{\bar{h}} + m(y) g'(y, h +
        \bar{h})\partial_{\bar{h}}  \right] \right\} \Bigg[ 1 - t(x, h_\eta
    + \bar{h})^2 \right. \nonumber \\ \fl &\left.  +
    \int_0^x d z\ \dot{ \chi}(z) m(z) \mbox{T}_y \exp\left\{\int_z^x d y\
      \dot{ \chi}(y)
      \left[\frac 12 \partial^2_{\bar{h}} + m(y) g'(y, h +
        \bar{h})\partial_{\bar{h}}  \right] \right\}\right. \nonumber \\
  &\qquad \left.  \qquad  \times (\partial_{\bar{h}} t(z, h_\eta +
    \bar{h}))^2 \Bigg]_{\bar{h}=0}\right\rangle_\eta \ge 0\ .
\end{eqnarray}
It was shown in Ref.~\cite{Janis07c} that equality in
Eq.~\eref{eq:stability-continuous} can be derived from a total
derivative of Eq.~\eref{eq:m-eq} with respect to variable $x$. It
means that if equation~\eref{eq:m-eq} is fulfilled for all variables
$x\in [0,1]$, both sides of Eq.~\eref{eq:stability-continuous} are
equal as well. We thereby proved that the Parisi continuous RSB
solution is marginally stable. The spin-glass susceptibility has
zero eigenvalue but no negative ones.

\subsection{Asymptotic solution near de Almeida-Thouless instability
  line}
\label{sec:continuous-asymptotic}

We now explicitly asymptotically solve the equations for the order
parameters of the continuous RSB scheme. We know from the discrete
version that $\dot{ \chi}(x) = 1$ for $x\in [0,X]$ and vanishes
elsewhere. We introduce a new dimensionless variable $\lambda = x/X$
that spans again interval $[0,1]$.  The physical parameter $X$ serves
as an expansion parameter in the asymptotic region near the AT
instability line.

In the continuous limit we have two basic equations to solve. It is
Eq.~\eref{eq:q-eq} for the SK order parameter $q$ and
Eq.~\eref{eq:m-eq} from which we determine $X$ and $m(\lambda)$. It
can be shown that for $x\le X$ equation~\eref{eq:chi-eq} is a total
derivative of Eq.~\eref{eq:m-eq}. There is thus no inconsistency if we
disregard Eq.~\eref{eq:chi-eq} as a stationarity equation for the free
energy of the continuous RSB solution.  We know from the preceding
subsection that the total derivative of Eq.~\eref{eq:m-eq} with
respect to $x$ expresses a marginal stability of the continuous RSB
state.

The asymptotic solution near the AT line is a polynomial in variable
$X$. We hence expand all quantities to a necessary order in this
parameter.  The $T$-exponential on the right-hand side of
Eq.~\eref{eq:m-eq} must be expanded to $X^3$. Further on, the
order-parameter function becomes also a polynomial in $X$. The
relevant order parameters are then expanded as follows
\numparts
\begin{eqnarray}\label{eq:q-expansion}
  q &= q_0 + X q_1' + X^2 q_2'\ ,\\ \label{eq:m-expansion}
  m(\lambda) & = m_0 + X\lambda\ m_1'  \
  . \end{eqnarray}\endnumparts

We first use Eq.~\eref{eq:q-eq} to simplify Eq.~\eref{eq:m-eq} and
then expand its right-hand side to $X^3$. Simultaneously we make use
of the expansion of the order parameters from
Eqs.~\eref{eq:q-expansion} and~\eref{eq:m-expansion}. We obtain a
cubic polynomial in $\lambda$.  Coefficients at each power of
$\lambda$ must vanish and we have three equations for parameters $m_0,
m_1'$ and $X$. Parameters $q_1'$ and $q_2'$ are determined from an
expansion of Eq.~\eref{eq:q-eq}. The explicit solution was calculated
with the aid of the programme MATHEMATICA and reads
\numparts
\begin{eqnarray}\label{eq:q0}
  q_0 & = \left\langle t_0^2\right\rangle_\eta\ ,\\
  \label{eq:q1}
  q_1'& = -  \ \frac {2 \beta^2(1 - m_0) \left\langle t_0^2(1 -
      t_0^2)\right\rangle_\eta} {1 - \beta^2\left\langle(1 -  t_0^2)(1 -
      3 t_0^2)\right\rangle_\eta}  \ , \\
  \label{eq:q2}
  q_2' & = -\  \frac {\left\langle Q_2\right\rangle_\eta}{1 -
    \beta^2\left\langle(1 -  t_0^2)(1 - 3 t_0^2\right\rangle_\eta}  \ .
\end{eqnarray}\endnumparts%
Here we denoted
\begin{eqnarray*}\fl
  Q_2 = \left( 1 - t_0^2\right) \left[\left(\beta ^2
      \left(m_0-1\right) \left(-7    t_0^2+\left(5 t_0^2-3\right)
        m_0+5\right)-m_1'\right) t_0^2\right. \nonumber \\ \left. + \left(15
      \left(t_0^2-1\right) t_0^2+2\right) \beta ^2 q_1^{\prime 2} -2
\left(10 t_0^4-9       t_0^2+1\right) \beta ^2 \left(m_0-1\right)
q_1'\right]\ . \end{eqnarray*} The above solution is then used in the
equations for $X, m_0, m_1'$. These three equations read
\numparts
\begin{eqnarray}\label{eq:L1}
  \fl
  1 &= \left\langle \left(1 - t_0^2\right)^2 \beta ^2 \left(1 - 2X
      \beta^2\left(m_0\left(2 t_0^2-\left(37 t_0^4 + 22 t_0^2 - 1\right)    X
          \beta ^2\right)  -X m_1' t_0^2\right. \right.\right. \nonumber \\ \fl  &
  \left. \left. \left. \qquad + \left(7 t_0^2-3\right)    X \beta ^2 m_0^2
        t_0^2\right)  +  X\beta^2 \left(10 t_0^2+\left(105 t_0^4 - 80
          t_0^2+7\right) X \beta ^2-2\right) \right)\right\rangle_\eta ,
  \\ \label{eq:L2}\fl 0 &= \left\langle \left(1 - t_0^2\right) \left(-2
      t_0^2-2 \left(21 t_0^4-14 t_0^2+1\right) X \beta^2 \right.\right.
  \nonumber \\ \fl & \left.\left. \qquad + m_0 \left(6 \left(t_0^2-1\right)
        X \beta ^2 m_0 t_0^2 +\left(-9 t_0^4+20
          t_0^2-3\right) X \beta ^2+1 - t_0^2\right) \right)\right\rangle_\eta \ , \\ %
  \fl 0 &= \left\langle \left(1 - t_0^2\right)\right. \nonumber \\
  \fl & \left. \qquad  \left(m_1'(1 - t_0^2) +2\beta^2 \left(1 -3 t_0^2
      \right)^2+3    \left(t_0^2-1\right) \beta ^2 m_0 \left(8
        t_0^2+\left(t_0^2-1\right)    m_0\right)\right)\right\rangle_\eta \ . %
  \label{eq:L3}
\end{eqnarray} \endnumparts
We determine $m_1'$ from Eq.~\eref{eq:L3}, $m_0$ then from
Eq.~\eref{eq:L2}. The two parameters we finally use in
Eq.~\eref{eq:L1} from which we calculate $X$. To find the leading
asymptotic behaviour of $X$ we have to expand it in powers of the
initial small parameter $\alpha$ measuring the depth of penetration
into the spin-glass phase. Parameter $\alpha$ was defined in
Eq.~\eref{eq:alpha-def} and emerges in Eq.~\eref{eq:L1} as an absolute,
$X$-independent term. The solutions for $X$ and $m_0$ from the
continuous RSB scheme then coincide in the leading order in $\alpha$
with the result for $ \chi_1$ and $m$ from the discrete $1$RSB
solution, Eqs.~\eref{eq:X-solution} and ~\eref{eq:m0-solution},
respectively. The last parameter $m_1'$ has an explicit representation
\begin{eqnarray}\label{eq:m1}
  m_1' =  -\ \frac {2\Delta m}{X}
\end{eqnarray}
where again parameter $\Delta m$ was already determined within the
discrete $2$RSB scheme in Eq.~\eref{eq:M-solution}. The
asymptotic limit of the full solution near the AT line is hence
completely determined by the parameters from the two-step RSB solution.

\section{Conclusions}
\label{sec:Conclusions}

We studied in this paper the behaviour of the replica-symmetry breaking
solutions in the discrete and continuous schemes. We started with the
discrete one with $K$ hierarchies and $2K + 1$ order parameters
$q,\Delta\chi_1, m_1,\ldots, \Delta\chi_K, m_K$. Equations for these order
parameters are derived from a local maximum of a free energy and were
explicitly solved in the asymptotic limit to the AT instability line. This
calculation served as an explicit manifestation of the way the Parisi
continuous RSB solution in the non-zero magnetic field is approached in
the limit $K\to\infty$. We found that unlike in zero magnetic field, the
SK solution is never, even asymptotically, stable in the non-zero field
below the AT line. The full solution in the low-temperature spin-glass
phase reduces near the AT line to the one-step RSB solution. We found that
$\sum_{l=1}^K\Delta\chi_l = \chi_1$ and $m_l$ do not depend in the leading
asymptotic order on the number of hierarchies $K$ and are exactly
determined by $1$RSB. Further on, we demonstrated that neither $\Delta m =
K(m_{l-1} - m_l)/2$  for $l \ge 2$ depends on the number of hierarchies
$K$ used. The characteristic parameters of the full asymptotic solution are
completely set by $2$RSB. There is no other parameter characterising the
asymptotic limit to the AT line. Increasing the number of hierarchies in
free energy~\eref {eq:mf-avfe} does not change the values $\chi_1=
\sum_{l=1}^K\Delta\chi_l $ and $\Delta m$. The new added order parameters
$ \chi_l, m_l$ for $l=3,\ldots, K$ are equidistantly distributed between
the edge values $0 \le \chi_l < \chi_1$ and $m_1 - \Delta m < m_l < m_1 +
\Delta m$, where $\chi_1$ and $m_1$ are calculated in $1$RSB. These
results explicitly prove that the discrete RSB scheme goes over in the
limit of infinite number of hierarchies to the Parisi continuous RSB
solution.

We analysed the behaviour of the discrete RSB scheme in the limit
$K\to\infty$ also generally. We performed this limit explicitly in the
stationarity equations maximising the free energy with finite-many
hierarchical levels. We derived in this way a set of equations for the
order parameters in the continuous limit. The equations for the order
parameters from the discrete RSB scheme goes over in the continuous
limit to two functional equations for order-parameter functions $\dot{
  \chi}(x)$ and $m(x)$. Since the former function comes up only under
integrals over the index variable $x\in[0,1]$, its point-wise
behaviour is irrelevant for the physical quantities. The only
significant information from $\dot{ \chi}(x)$ is an integral $\int_0^1d x
\dot{ \chi}(x) = X$. We hence can disregard the defining equations for
$\dot{ \chi}(x)$ and take explicitly into account only equations for the
SK parameter $q$ and for function $m(x)$.  Parameter $X$ is determined
from a combination of the two equations. We do not lose any
information by neglecting the defining equation for $\dot{ \chi}(x)$
that was shown to be a total derivative of the equation for
$m(x)$. This feature expresses a degeneracy of the stationarity
equations in the discrete RSB scheme. The equations resulting from the
continuous limit of the equations from the discrete scheme are
identical with those derived directly from the Parisi free energy via
a saddle point in Ref.~\cite{Janis07c}.  It means that the local
maximum of the Parisi free energy is a limit of local maxima of
discrete hierarchical free energies~\eref {eq:mf-avfe} when
$K\to\infty$. The continuous limit is analytical and all physical
quantities can be defined and calculated either directly from the
Parisi free-energy functional of Ref.~\cite{Janis07c} or from the
limit $K\to\infty$ of quantities introduced in the discrete scheme
with free energy~\eref{eq:mf-avfe}.

The explicit asymptotic solution of the discrete $K$RSB scheme enabled
the calculation of its thermodynamic stability. We found that in the
leading asymptotic order of the discrete scheme all the stability
conditions~\eref{eq:AT-hierarchical} and~\eref{eq:AT-zero}
coincide. Their value is negative for any finite number of hierarchies
$K$ but approaches zero as $K^{-2}$. The continuous scheme is then
marginally stable with no negative eigenvalue of the spin-glass
susceptibility. The SK replica-symmetric solution is asymptotically
stable in the leading order below the critical temperature only in
zero magnetic field. In the non-zero magnetic field the spin-glass
state goes over asymptotically to the one-step
replica-symmetry-breaking solution ($K=1$) that is marginally stable
in the leading order near the AT instability line.
 
\section*{Acknowledgement}

Research on this problem was carried out within a project AVOZ10100520
of the Academy of Sciences of the Czech Republic.

\end{document}